\documentclass[aps,preprint]{revtex4}
\usepackage{epsfig}

\begin{document}

\title{Contact-induced spin polarization in carbon nanotubes} 

\author{Mauro S. Ferreira and Stefano~Sanvito} 
\affiliation{Physics Department, Trinity College, Dublin 2, Ireland}

\date{\today}

\begin{abstract}
Motivated by the possibility of combining spintronics
with molecular structures, we investigate the conditions for the
appearance of spin-polarization in low-dimensional tubular systems by
contacting them to a magnetic substrate. We derive a set of general
expressions describing the charge transfer between the tube and the
substrate and the relative energy costs. The mean-field solution of
the general expressions provides an insightful formula for the induced
spin-polarization. Using a tight-binding model for the electronic
structure we are able to estimate the magnitude and the stability of
the induced moment. This indicates that a significant magnetic moment
in carbon nanotubes can be observed.
\end{abstract}

\maketitle

\section{Introduction} 
Over the last decade there has been an explosive increase of activity
in two key areas of material science: spintronics and molecular
electronics. Spintronics is based on the use of the spin degrees of
freedom as well as the electronic charge for a number of
applications \cite{wolf}. The field has expanded significantly since
the discovery of the giant magnetoresistance effect in magnetic
multilayers \cite{Fert} and has been the main driving force leading to
the development of the present generation of magnetic storage devices.

Although the field has already demonstrated part of its potential, it
is worth noting that most of the proposed applications simply
translate well known concepts of conventional electronics into spin
systems. The typical devices are made with molecular beam epitaxy
growth, and lithographic techniques; a bottom-up approach to
spintronics devices has scarcely been explored. In this respect
molecular electronics provides the opposite approach \cite{molt}. Here
the basic idea is to use molecular systems for electronic applications
and conventional electronic devices such as transistors \cite{mol1},
negative differential resistors \cite{mol2} and rectifiers \cite{mol3}
have already been produced at the molecular level.

A few experiments have attempted to combine spintronics with
molecular devices. In their pioneering experiment \cite{Bruce}
Tsukagoshi and coworkers demonstrated that the $I$-$V$ curve of a
carbon nanotube sandwiched between two Co contacts presents hysteresis
when a magnetic field is applied. Such spin-valve behavior indicates
spin injection into the nanotube with a spin diffusion length (the
average distance that an electron travels before flipping its spin
direction) of the order of 100 nm. This makes carbon nanotubes very
attractive for spintronics applications. Other carbon structures are
capable of accommodating net spin polarization and Coey {\it et al}
have shown evidence for a strong induced magnetic polarization at room
temperature in a graphite system with embedded ferromagnetic
nanoclusters \cite{mike}.

From these experiments it emerges that, on the one hand spins can
propagate in carbon materials almost without flipping their direction,
and on the other that the proximity with magnetic materials can induce
spin polarization in graphite-based systems. Although more controlled
experiments on synthetic nanostructures are needed, we believe that
the implementation of spin physics in carbon systems is possible and 
it will be crucial for the development of smaller and more sophisticated 
magneto-transport devices.

Motivated by the idea of combining spintronics with molecular
structures, we investigate the conditions for which an induced spin
polarization appears in a low-dimensional tubular molecule contacted
to a magnetic material. Although carbon nanotubes are the immediate
motivation for this work, our formalism is rather general and can be
applied to any cylindrical-like structures.

This paper is organized as follows. In the next section, we derive a
general expression for the charge transfer that occurs when a tubular
molecule is side-contacted to a metallic substrate. A complementary
expression for the contact-induced total energy change is also
presented. This set of expressions determines not only how the charge
is redistributed when the tubular molecule is contacted to the
substrate but also provides information about the stability of the
transfer process. The expressions derived in section \ref{sec2} do not
present any explicit spin dependence. In section \ref{sec3} we
generalize them to include the spin asymmetry of magnetic substrates
and we demonstrate that a net magnetic moment can be induced in the
tubular molecule. An estimation of the magnitude of the induced magnetic
moment and of its stability is then given.

\section{Contact-induced charge transfer and energy gain}
\label{sec2}

In order to investigate how a magnetic contact affects the spin
polarization of a nanotube, we start by calculating the change in the
electronic structure of a tubular molecule side-contacted to a
substrate. We model the contact by introducing an electronic coupling
between the tube and the substrate that accounts for the possible charge
transfer between the two materials. The interaction is assumed to be
only between the two lines of atoms, one on the tube and one on the
substrate, that are in closest proximity. Spin-dependent charge
transfer is likely to arise due to the spin asymmetry of the magnetic
substrate, thus leading to a net induced spin polarization.

The spin-dependent density of states is the relevant quantity to
calculate and provides the necessary information about whether or not
the contact leads to a net polarization of the nanostructure. We start
by making no assumptions regarding the specific models describing the
electronic structure of the system. In this way, we express the
spin-polarized density of states in terms of single-particle Green
functions matrix elements that can be calculated by different
techniques based on model Hamiltonians. Such a model-independent
treatment emphasizes the generality of our results and leads to a set
of closed-form expressions that provide a general method to
investigate charge transfer between contacting materials.

We consider infinitely long tubes of diameter $d$, which can be
thought of as 2-dimensional finite-width stripes wrapped around in
cylindrical shape. We also assume that N atoms are placed along the
circumference. The tubular system shows translational symmetry along
the longitudinal direction. Therefore, the electronic states along
this direction are well described by a reciprocal-space wave vector
that runs within the 1-dimensional Brillouin zone. Since the
translational symmetry is broken by the line of atoms contacting the
substrate, it is convenient to use real space coordinates along the
circumferential direction. In this way, electronic states are
identified by the pair of indexes $(j,k)$, where $k$ corresponds to
the wave vector along the longitudinal direction and $j$ labels the $N$
lines of atoms on the tube surface. Since the substrate also has
translational symmetry along the axial direction of the tube, the
electronic states of the substrate can be labeled by the same pair of
indexes. 

In terms of the single-particle Green function, the total density of
states $\rho(E)$ is written as
\begin{equation}
\rho(E)= (-{1 \over \pi})\, Im \, \sum_k \sum_j G_{j,j}(E,k)\,\,\,,
\label{dos}
\end{equation}
where $G_{j,j}(E,k)$ is the Green function of an electron with energy
$E$ moving on line $j$ with wave vector $k$. The sum over $j$ accounts
for all the atomic lines of the tube and the substrate.

It is convenient to define another Green function associated with the
disconnected system, that is, the isolated substrate and tube. In this
case, the translational symmetry along the circumferential direction
is reestablished and the electronic states on the tube are usually
described by a set of two wave vectors; the longitudinal component
running continuously over the 1-dimensional Brillouin zone and a
finite set of quantized wave vectors induced by the cylindrical
boundary conditions. Likewise, the in-plane translational symmetry is
also restored for the substrate. However, to be consistent with the
notation in terms of the indexes $(j,k)$, we label the Green function
of the disconnected system as ${\cal G}_{j,j^\prime}(E,k)$ describing
electronic propagation between lines $j$ and $j^\prime$. To
distinguish between atomic lines on the tube and on the substrate, we label
the former by integers $j \leq 0$ and the latter by $j \geq 1$. It is
clear that ${\cal G}_{j,j^\prime}$ vanishes if $j$ and $j^\prime$
refer to lines on different subsystems.

The effect of the contact on the total density of states can be
calculated by summing up the corresponding change in the Green
function over all possible lines, {\it i.e.}, {$\sum_{j}\Delta
G_{j,j}$, where $\Delta G_{j,j} = G_{j,j} - {\cal G}_{j,j}$ gives the
variation of the Green function evaluated at line $j$. Consistently
with our notation, we define lines $j=0$ and $j=1$ as those atomic
lines connecting the tube with the substrate, respectively. Therefore,
the matrix elements of $\Delta G_{j,j}$ are
\begin{equation}
\Delta G_{j,j} = {\cal G}_{j,0} \, t \, (1 - {\cal G}_{1,1} \, t^\dag
\, {\cal G}_{0,0} \, t)^{-1} \, {\cal G}_{1,1} \, t^\dag \, 
{\cal G}_{0,j}
\end{equation}
and
\begin{equation}
\Delta G_{j,j} = {\cal G}_{j,1} \, t^\dag \, (1 - {\cal G}_{0,0} \, t
\, {\cal G}_{1,1} \, t^\dag)^{-1} \, {\cal G}_{0,0} \, t \, {\cal
G}_{1,j}\;,
\end{equation}
for lines on the tube ($j \leq 0$) and on the substrate ($j \geq 0$),
respectively.  In the equations above, we have introduced the
parameter $t$ describing the coupling between the tube and the
substrate. This quantity plays the role of a tight-binding-like
energy-independent electronic hopping between the relevant overlapping
orbitals on either side. The Green functions above are energy- and
$k$-dependent matrices, whose indexes may refer to orbital and spin
degrees of freedom.

We define $\Delta \rho(E) = \rho(E) - \rho_0(E)$ as the density of
states change between the disconnected ($\rho_0$) and the side-contacted
system ($\rho$). According to Eq.(\ref{dos}) the variation of the
total density of states is given by
\begin{equation}
\Delta \rho(E)= (-{1 \over \pi})\, Im \, Tr \, \sum_k \sum_j \Delta 
G_{j,j}(E,k)\,\,\,,
\end{equation}
where the trace accounts for possible internal degrees of freedom such 
as spin and orbital indexes. By combining the cyclic property of the
trace with the definition of Green functions, we can write $\Delta
\rho(E)$ as 
\begin{equation} 
\Delta \rho(E)= -({1 \over \pi})\, Im \, Tr \, \sum_k (1 - {\cal
G}_{0,0} t {\cal G}_{1,1} t^\dag)^{-1} [ {d{\cal G}_{0,0} \over dE} \,
t \, {\cal G}_{1,1} \, t^\dag \, + \, {\cal G}_{0,0} \, t \, {d{\cal
G}_{1,1} \over dE} \, t^\dag ]\,\,,
\end{equation}
or in a further simplified form
\begin{equation}
\Delta \rho(E)= - ({1 \over \pi})\, Im \, \sum_k {d \over dE} 
\ln \det(1 - {\cal G}_{0,0} \, t \, {\cal G}_{1,1} \, t^\dag)\,.
\label{The Dyson}
\end{equation}
Equation (\ref{The Dyson}) writes $\Delta \rho(E)$ in terms of the
Green function matrix elements for the disconnected system (${\cal
G}$) plus the coupling parameters $t$ and $t^\dag$ between the tube
and substrate. More specifically, it only depends on the diagonal
elements ${\cal G}_{0,0}$ and ${\cal G}_{1,1}$, namely those where the
connection takes place. Eq.(\ref{The Dyson}) is therefore a convenient
expression to calculate the effect of the coupling since it provides
the required change in the density of states without the need of
evaluating the electronic structure for the connected and disconnected
system separately. This is equivalent to the Lloyd's formula
\cite{Lloyd} describing the variation of density of states due to a
diagonal perturbation, a very useful method to treat substitutional
impurities. This method has also been used in the study of magnetic
coupling between impurities in metallic systems \cite{coupling}. The
fundamental difference in the case presented here is that the
perturbation has an off-diagonal form representing the contact between
the two structures. It is worth highlighting that the expression above
involves no approximation and is exact for arbitrary values of the
parameters $t$ and $t^\dag$.

From the variation in the total density of states, we can derive the
change in the number of electrons $\Delta N$ (at zero
temperature). This is obtained by integrating $\Delta \rho$ up to the
Fermi level $E_F$, {\it i.e.},
\begin{equation} 
\Delta N(E_F)= - ({1 \over \pi}) \, \sum_k Im \, \ln \, \det(1 - {\cal
G}_{0,0}(E_F) \, t \, {\cal G}_{1,1}(E_F) \, t^\dag)\,\,.
\label{DN}
\end{equation}
In the equation above the relevant matrix elements are evaluated at
the Fermi level. Since $\Delta \rho(E)$ is the variation of
the total density of states, its integral gives the change of number
of particles in the closed system. This is of course a conserved quantity
and the equation $\Delta N(E_F) = 0$ allows us to calculate the value of the
Fermi level.

Another quantity derivable from the change in the density of states is
the effect of the contact on the total electronic energy. This is a
fundamental quantity whose value determines whether or not the
perturbation in the electronic structure is energetically
favorable. It is defined as
\begin{equation}
\Delta E = \int_{-\infty}^{E_F} dE \, E \, \Delta \rho(E)\,\,.
\end{equation}
From the expression for $\Delta \rho(E)$ in Eq.(\ref{The Dyson}) we
have that 
\begin{equation} 
\Delta E = ({1 \over \pi}) \, \sum_k \int_{-\infty}^{E_F} dE \, Im \,
\ln \, \det(1 - {\cal G}_{0,0} \, t \, {\cal G}_{1,1} \, t^\dag)\,\,.
\label{DE}
\end{equation}
If $\Delta E < 0$, the changes in the electronic structure predicted
by the equations (\ref{The Dyson}), (\ref{DN}) and (\ref{DE}) are
possible when the energy gain is sufficient to overcome the energy costs
involved in the transition.

The changes $\Delta \rho$, $\Delta N$ and $\Delta E$ describe the
effect of the contact on the density of states, number of particles,
and total energy of the entire system, {\it i.e.}, the tube and the
substrate. In order to investigate possible contact-induced
spin-polarizations, one must look at similar changes on the separate
parts. In other words, instead of evaluating the total density of
states summed over all possible sites in the structure, we must
distinguish between the changes in the tube and in the substrate.
Bearing in mind that global charge neutrality is imposed by
Eq.(\ref{DN}), any modification in the total number of particles on
the tube must be compensated by the corresponding change on the
substrate. Therefore, to calculate the charge transfer between the
tube and substrate it is sufficient to evaluate the variation of
number of particles on either part. We choose to focus on the tube and
calculate the change of density of states summed over all atomic lines
of the tube. Analogously to the derivation presented above, the change
in the density of states ($\rho_t$) on the tube is given by
\begin{equation}
\Delta \rho_t(E)= ({1 \over \pi})\, Im \, Tr \, \sum_k (1 - {\cal
G}_{1,1} \, t^\dag \, {\cal G}_{0,0} \, t)^{-1} \, {d {\cal G}_{0,0}
\over dE} \, t \, {\cal G}_{1,1} \, t^\dag\,\,.
\label{Drhot}
\end{equation}
The charge transfer to the tube ($\Delta N_t$) is the integral of the
above expression and is written as
\begin{equation}
\Delta N_t(E)= ({1 \over \pi})\, Im \, Tr \, \sum_k
\int_{-\infty}^{E_F}\,dE\,(1 - {\cal G}_{1,1} \, t^\dag \, {\cal G}_{0,0} \,
t)^{-1} \, {d {\cal G}_{0,0} \over dE} \, t \, {\cal G}_{1,1} \,
t^\dag\,\,.
\label{DNt}
\end{equation}

Equations (\ref{DE}) and (\ref{DNt}) are the fundamental results of
this work. The first tells us whether or not the charge transfer is
energetically favorable, and the second the amount of charge exchanged
between the tube and the substrate.
These form a closed system of equations
written in terms of the Green functions of the disconnected system and
the coupling parameters. It is important to stress that although the
complexity involved in evaluating the expressions for $\Delta N_t$ and
$\Delta E$ depends on the choice of the Hamiltonian used to describe
the electronic structure of the system, the validity of equations
(\ref{DE}) and (\ref{DNt}) does not. This means that our expressions
can be equally used with simple model Hamiltonians or with a full
realistic description of the electronic structure.

Equation (\ref{Drhot}) for $\Delta\rho_t$ is not as concise as its
counterpart Eq.(\ref{The Dyson}) but it can be further simplified by
expanding it to second order in $t$. This approximation is valid in
the limit of weak coupling, which is satisfied in the case of carbon
nanotubes sitting on top of transition metals. In fact, recent density
functional theory calculations of graphite on a [001] cobalt surface
suggests a value for the coupling parameter of $t = W/30$, where $W$
is the width of the graphite $\pi$-band \cite{Goto}. Furthermore, when
the sum over $k$ is eliminated and $\Delta \rho_t(E)$ is integrated up
to the Fermi level, $\Delta N_t(E_F)$ becomes
\begin{equation} 
\Delta N_t(E_F) = Tr\, \int_{-\infty}^{E_F} dE \,\,\left[ \rho_0(E) {d
V_1(E) \over dE} + \rho_1(E) {d V_0(E) \over dE} \right]\,\,,
\label{DNt2}
\end{equation}
where $\rho_0(E)$ and $\rho_1(E)$ are the density of states on the
tube and on the substrate, respectively. $V_m(E)= t Re[{\cal G}_{m,m}]
t^\dag$ ($m = 0,1$) plays the role of an energy-dependent electronic
potential. $V_1(E)$ is the potential felt by the tube due to the
substrate and $V_0(E)$ is the analogous potential felt by the
substrate and produced by the tube. It is worth noting that the
potentials $V_0(E)$ and $V_1(E)$ depend on the real part of the Green
functions ${\cal G}_{0,0}$ and ${\cal G}_{1,1}$, respectively. These
two quantities are directly obtainable from electronic structure
calculations for the disconnected system. A similar second-order
expansion of Eq.(\ref{DE}) also simplifies the formula for the energy
change $\Delta E$, which now reads as
\begin{equation} 
\Delta E = Tr \, \int_{-\infty}^{E_F} dE \, \left[ \rho_0(E) V_1(E) +
\rho_1(E) V_0(E) \right] \,\,.
\label{DE2}
\end{equation}

Although Eqs. (\ref{DNt2}) and (\ref{DE2}) represent a more concise
version of their respective counterparts, Eqs. (\ref{DNt}) and
(\ref{DE}), they are still in integral forms. These can be further
simplified by replacing $V_m(E)$ with its mean value $\langle V_m(E)
\rangle$. This approximation gives rise to the following two
expressions:
\begin{equation} 
\Delta N_t(E_F) = Tr \left\{ \, \rho_0(E_F) \left[ V_1(E_F) - \langle V_1
\rangle \right] + \rho_1(E_F) \left[ V_0(E_F) - \langle V_0 \rangle
\right] \, \right\}
\label{DNmf}
\end{equation}
and 
\begin{equation} 
\Delta E = Tr \left\{ N_0(E_F) \langle V_1 \rangle + 
N_1(E_F) \langle V_0 \rangle \right\} \,\,.
\label{DEmf}
\end{equation} 
In this form, the expressions for the charge transfer and the
respective energy gain are written in terms of the density of states
for both the tube and the substrate, and the potentials
$V_0(E_F)$ and $V_1(E_F)$, all evaluated at the Fermi level.  They
also depend on the total number of electrons $N_0$ and $N_1$.

In calculating the average $\langle V_0 \rangle$ ($\langle V_1
\rangle$), the integration limits are not in the range $[-\infty,E_F]$
as in Eqs.(\ref{DNt2}) and (\ref{DE2}) but start from the bottom of
the electronic band $\rho_1$ ($\rho_0$). The upper integration limit
is common to both cases and is given by the Fermi level $E_F$. It is
clear from Eq.(\ref{DNmf}) that the sign of $\Delta N_t(E_F)$ is fully
determined by the potentials $V_0$ and $V_1$. The side-contacted
nanotube will then be electron- (hole-) doped for positive (negative)
values of $\Delta N_t(E_F)$. This means that the balance between
$\langle V_m \rangle$ and $V_m(E_F)$ determines the type of charge
transfer between the structures. We have checked the results obtained
by the mean-field equations against those predicted by Eqs.(\ref{DNt})
and (\ref{DE}), and we find both a qualitative agreement and values of
the same order of magnitude.

\section{Magnetic substrates}
\label{sec3}

The expressions presented in the previous section display no explicit
dependence of the electronic structure on the spin degree of freedom.
However, when a magnetic substrate is considered the spin symmetry is 
broken. In this case the expressions derived in the previous section are
still valid since an explicit spin-dependence can be added to both the 
Green functions and the coupling parameter, without loss of generality.
Although general non-collinear spin Hamiltonians can be considered, we 
restrict our analysis to collinear spin in the two-spin fluid model.
Within this model all the quantities are diagonal in the spin subspace and 
the only variation with the spin-degenerate case 
is that the expressions for the charge transfer and the energy gain are
different for the two spin sub-bands. Therefore any induced magnetization 
$M$ on the tube results from the spin-imbalance of the charge transfer, 
$M = ( \Delta N_t^\uparrow - \Delta N_t^\downarrow ) \mu_B$, where $\Delta
N_t^\sigma$ is the charge transfer for a spin $\sigma$ and $\mu_B$ is
the Bohr magneton. When the substrate is magnetic 
the charge transfer for the majority-spin sub-band
is different from that of the minority, leading to a net
induced magnetic moment on the tube.

It is worth recalling that equations (\ref{DNt2}) and (\ref{DE2}) are
complementary and that charge will be transferred only if the
corresponding energy gain is sufficient to outweigh the energy
costs. In the two-spin fluid model we have to calculate the energetics
of the charge transfer process for each spin direction. Only when
the energy gain is favorable for both spins does the quantity $(\Delta
N_t^\uparrow - \Delta N_t^\downarrow)\mu_B$ describe the induced
moment. In other words, if the energy gain for one spin-direction is
not sufficient to surpass the energy costs, the corresponding charge
transfer will not take place and the difference $M = ( \Delta
N_t^\uparrow - \Delta N_t^\downarrow ) \mu_B$ will be
meaningless. Assuming that both transitions are energetically
favorable we can make use of equation (\ref{DNt2}) to write the
induced moment $M$ as
\begin{equation} 
M = \mu_B \, Tr\, \int_{-\infty}^{E_F} dE \,\,\left\{ \rho_0(E) {d
\over dE} [ V_1^\uparrow(E) - V_1^\downarrow(E) ] + [
\rho^\uparrow_1(E) - \rho^\downarrow_1(E) ] {d V_0(E) \over dE}
\right\}\,\,,
\label{M}
\end{equation}
where the spin-polarization $\sigma$ is now explicitly included in the
quantities describing the substrate. Bearing in mind that the
spin-bands are split by the exchange integral $\Delta$ and neglecting
possible hybridization effects, we can approximately correlate the
majority- and minority-spin bands by $V_1^\downarrow(E) =
V_1^\uparrow(E - \Delta)$ and $\rho_1^\downarrow(E) =
\rho_1^\uparrow(E - \Delta)$.  A further simplification can be made by
expanding the substrate quantities in powers of $\Delta$. In this
case, the induced magnetization becomes
\begin{equation} 
M = - \mu_B\, \Delta \,\,\, Tr\, \int_{-\infty}^{E_F} dE \,\,\left\{
\rho_0(E) \, {d^2 V_1^\uparrow(E) \over dE^2} + {d \rho^\uparrow_1(E)
\over dE} \, {d V_0(E) \over dE} \right\}\,\,.
\label{M1}
\end{equation}
Both equations (\ref{M}) and (\ref{M1}) give the induced magnetic
moment in terms of quantities that are directly obtainable from
electronic structure calculations and provide valuable expressions to
determine the contact-induced spin polarizations. Whereas the latter
is valid for magnetic substrates whose spin bands are not
significantly split, the former gives a general expression for the
induced moment with no limitations about the electronic structure
parameters.

\begin{figure}[ht]
\epsfxsize=9cm
\centerline{\epsffile{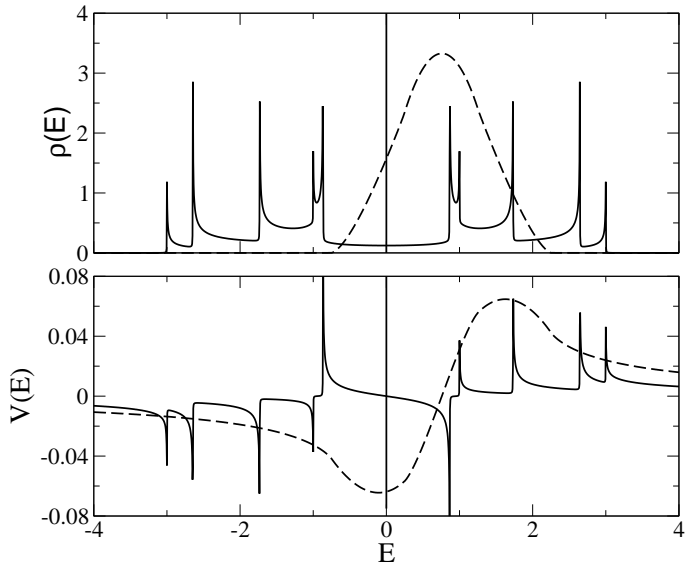}}
\caption{Density of states $\rho_0$ and $\rho_1$ and potentials $V_0$
and $V_1$ for a typical case: a (3,3) carbon nanotube (solid line)
attached to a semi-infinite cubic slab (dashed line). The electronic
structure parameters for the slab leads to a band that is $5 \, eV$
wide and centered at $\epsilon_1 = 0.75 \gamma$. The Fermi level is
fixed at $E_F= 0$ (vertical line).}
\label{F1}
\end{figure}
In order to test whether or not the magnetic contact can induce
spin-polarization on the tube we must determine the order of magnitude
of the charge transfer for a given spin band. This can be done within
a simplified model that contains the fundamental features of the
electronic structure of both the nanotube and the substrate. The
single-band tight-binding model is known to reproduce well the band
structure of both graphite and nanotubes of somewhat large
diameters. The electronic hopping within the tube is described by the
parameter $\gamma = 2.5 \, eV$ and is hereafter used as our energy
unit. Likewise, the d-band of magnetic transition metals can be
described within the same model by an appropriate choice of band width
and total number of electrons. Figure 1 shows the density of states
$\rho_0$ and $\rho_1$ as well as the potentials $V_0$ and $V_1$ for a
typical case. We have chosen an armchair nanotube with $N = 12$ atoms
per ring. The substrate is modeled by a semi-infinite cubic slab whose
electronic structure parameters lead to a band that is $5 \, eV$ wide
(typical of transition metals) and that is centered at an arbitrary
position $\epsilon_1 = 0.75 \gamma$. The Fermi level is fixed at $E_F
= 0$ and is represented in the figure by a vertical line. For
this choice of parameters, the calculated results are a charge
transfer of $\Delta N_t = -2 \times 10^2$ electrons/$\mu m$ with the
respective energy gain of $\Delta E = - 6.5 \times 10^{-2} \,eV$ per
unit cell. It is worth recalling that the present calculations are for
zero temperature, although the finite-temperature regime can in
principle be accounted for by including the Fermi functions in the
integrands of the expressions above. The negative sign of $\Delta N_t$
indicates that for this particular band alignment, electrons flow from
the nanotube onto the substrate.

The specific values of $\Delta N_t$ and $\Delta E$ depend on the
particular alignment of the electronic bands of the tube and the
substrate. We investigate different possibilities by changing the
on-site potential $\epsilon_1$ of the substrate atoms. This corresponds to
shifting the band center along the nanotube energy spectrum. In figure 2
we show the changes $\Delta N_t$ and $\Delta
E$ as a function of $\epsilon_1$ for two different tube
diameters. Since the Fermi energy is kept at $E_F = 0$, a shift in the
electronic band of the substrate also affects the total number of
electrons $N_1$ on the substrate. 
\begin{figure}[ht]
\epsfxsize=9cm
\centerline{\epsffile{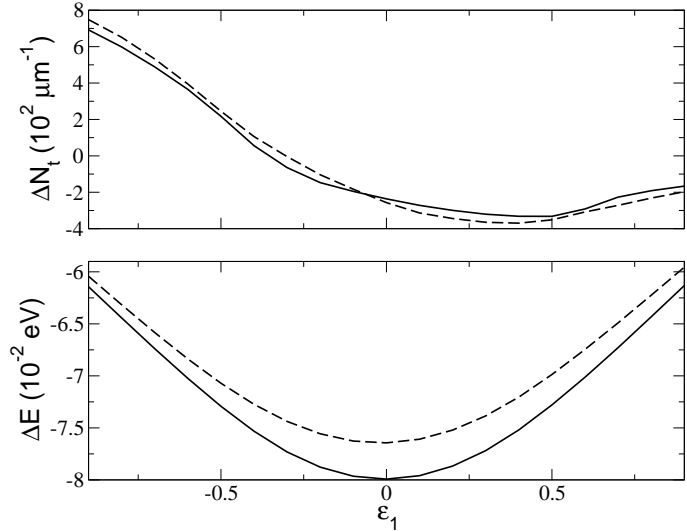}}
\caption{Charge transfer ($\Delta N_t$) and the respective energy gain
per unit cell ($\Delta E$) for different band alignments. The
parameter $\epsilon_1$ corresponds to the center of the substrate
band. Solid and dashed lines refer to (3,3) and (8,8) armchair
nanotubes, respectively.} 
\label{F2}
\end{figure}
The figure shows that the charge transfer $\Delta N_t$ can change
sign, indicating that the tube can be doped either with electrons or
with holes. However, despite differences in sign, the magnitude of the
charge transfer $\vert \Delta N_t \vert$ does not change substantially
and reaches values up to $8 \times 10^2$ electrons/$\mu m$. The fact
that the charge transfer between the tube and the substrate can change
sign depending on the band alignment, has important consequences on
the induced magnetic moment. In fact, if the band splitting of the
substrate is such that the charge transfer for the majority-spin
sub-band has opposite sign to that of the minority-spin band, the spin
balance on the nanotube is not only broken but maximized. In this case
electrons of opposite spins flow in opposite directions (for instance,
majority spins will flow from the substrate into the tube, and
minority from the tube into the substrate). The picture shows that
this is the case when the Fermi level lies close to opposite edges of
the ferromagnetic spin bands. Half metals seem to satisfy this
requirement and therefore are potential candidates for inducing large
spin imbalance in nanotubes. In other words, magnetic substrates made
of half metals are predicted to be the best materials to induce a
magnetic moment in a nanotube. Considering the results of figure 2 as
a reference, we estimate that the maximum value of induced
magnetization is $M = 10^{-1} \, \mu_B$ per unit cell, a magnitude
that is experimentally detectable. The stability of this induced
moment can also be estimated by the energy gain $\Delta E$, which is
in the order of $10^{-1} eV$, as shown in figure 2. Induced moments in
tubes of increasingly large diameters are less stable and $\Delta E$
must saturate toward the value associated with a nanotube deposited on
a graphite substrate. Furthermore, we note that the lattice relaxation
\cite{capaz03} induced by the charge transfer as well as the charging
energy due to the low capacitance of carbon nanotubes
\cite{capacitance} must be considered when calculating the total
energy costs of the electronic transition. Although these two factors
are influential on whether the transfer process becomes energetically
favorable, they are at least one order of magnitude below the energy
gain obtained by the change in the electronic structure. In fact, the
typical value for the quantum capacitance\cite{capacitance2} (per unit
cell) of a nanotube is $C = 0.35 \, e/V$. For the parameters used in
Figure \ref{F1}, this leads to a charging energy $E_c = 0.0025 \, eV$,
which is still substantially smaller that the energy gain of $\Delta E
= 0.065 \, eV$ calculated here.

Finally, we wish to briefly comment on the similarities between our
contact-induced spin-polarization effect and the problem of
spin-injection from metallic systems. Whereas the latter is a
non-equilibrium transport effect, the former is the result of charge
and spin rearrangement toward the equilibrium configuration between
the magnetic and non-magnetic materials. In other words, while in the
spin-injection problem the electrons must travel long distances to be
probed by a detector, this is not the case for the contact-induced
spin-polarization. The two phenomena can indeed be addressed by a
common formalism but a complete comparison would require a
reformulation of our method in terms of transport quantities, which is
beyond the scope of the present paper.

\section{Conclusions}

In summary, we have presented a model that describes the charge 
transfer of a carbon nanotube in contact to a substrate. 
Closed-form expressions in terms of Green functions give
the charge transfer and the respective energy gain associated with the
transition. We have subsequently shown that when the substrate is
magnetic the spin imbalance of the surface may lead to an induced spin
polarization in the nanotube. Within a simple model that reproduces the basic
features of the electronic structures of both nanotubes and transition
metal surfaces, we were able to estimate the magnitude and the
stability of the induced moment. This indicates that a measurable
magnetic moment can be induced in carbon nanotubes when contacted to
magnetic substrates. Finally, we have demonstrated that half metals
are the best candidates for inducing a sizable magnetic moment in
carbon nanotubes.

\section*{Acknowledgments}

The authors are grateful to C.J. Lambert, J.M.D. Coey, and R.B. Muniz
for helpful discussions. SS thanks Enterprise Ireland (grant EI-SC/2002-10)
for the financial support.

\end{document}